\RequirePackage{lineno}
\documentclass[linenumbers,author-year,preprint,showkeys,preprintnumbers,amsmath,amssymb,superscriptaddress]{revtex4}

\usepackage{graphicx}
\usepackage{graphics}
\usepackage{subfigure}
\usepackage{dcolumn}
\usepackage{bm}
\usepackage{txfonts}
\usepackage{enumerate}
\usepackage{amsmath,empheq,mathrsfs}

\begin{document}


\centerline{Phenotype affinity mediated interactions can facilitate the evolution of cooperation}

{\small {\vskip 12pt \centerline{Te Wu$^{1}$, Feng Fu$^{2}$, Long Wang$^{3}$}

\begin{center}
$^1$ Center for Complex Systems, Xidian University, Xi'an, China \\
$^2$ Department of Mathematics, Dartmouth College, Hanover, United States of America\\
$^3$ Center for Systems and Control, College of Engineering, Peking University, Beijing, China
\end{center}
}}

%
%
%

\begin{center}
{
\begin{minipage}{142mm}
{\bf Abstract:} \, We study the coevolutionary dynamics of the diversity of phenotype expression and the evolution of cooperation in the Prisoner's Dilemma game. Rather than pre-assigning zero-or-one interaction rate, we diversify the rate of interaction by associating it with the phenotypes shared in common. Individuals each carry a set of potentially expressible phenotypes and expresses a certain number of phenotypes at a cost proportional to the number. The number of expressed phenotypes and thus the rate of interaction is an evolvable trait. Our results show that nonnegligible cost of expressing phenotypes restrains phenotype expression, and the evolutionary race mainly proceeds on between cooperative strains and defective strains who express a very few phenotypes. It pays for cooperative strains to express a very few phenotypes. Though such a low level of expression weakens reciprocity between cooperative strains, it decelerates rate of interaction between cooperative strains and defective strains to a larger degree, leading to the predominance of cooperative strains over defective strains. We also find that evolved diversity of phenotype expression can occasionally destabilize due to the invasion of defective mutants, implying that cooperation and diversity of phenotype expression can mutually reinforce each other. Therefore, our results provide new insights into better understanding the coevolution of cooperation and the diversity of phenotype expression.

\vspace*{1\baselineskip}

\end{minipage}
}
\end{center}

\clearpage


\section{Introduction}

The emergence and persistence of cooperative behaviors in populations of selfish individuals is a key problem in evolutionary biology~\cite{alexrod(1981)science, rakoff-nahoum(2016)nature, strassmann(2008)nature, taylor(2007)nature, wedekind(2000)science, hauser(2014)nature, imhof(2010)prsb, vukov(2005)pre, mcnamara(2004)nature, szolnoki(2015)prsb, li(2017)science}. The Prisoner's Dilemma game provides a convenient paradigm to address this problem. In the game, a cooperator brings a benefit of $b$ to her coplayer at a cost of $c$ to herself. A defector produces no benefit and incurs no cost. As $b>c>0$, the best strategy for an individual is to defect irrespective of her coplayer's strategy. However, the total payoff would be maximized if both have adhered to cooperation. The dissonance in the optimal strategy for an individual and for the group leads to the social dilemma.

To resolve the dilemma, a variety of mechanisms~\cite{nowak(2006)science, masuda(2007)prsb} promoting the evolution of cooperation have been proposed. The tag based interactions have been intensively studied~\cite{riolo(2001)nature, masuda(2015)pr, roberts(2002)nature, riolo(2002)nature, tanimoto(2007)jtb, traulsen(2003)pre, traulsen(2004)pre, traulsen(2007)plosone, wute(2013)jtb, fu(2012)sr}. Its essence lies in that individuals can recognize each other by tags and cooperators just help those sufficiently similar to themselves. By this simple rule, the authors~\cite{riolo(2001)nature} have pointed out that cooperation can arise without reciprocity. When ``never to donate" is incorporated as a possible strategy, similarity can still breed cooperation but only when mutations towards ``never to donate" are not very strong~\cite{roberts(2002)nature, riolo(2002)nature}. Traulsen and his coworkers have captured the essence of this model~\cite{riolo(2001)nature} by considering two tags and two levels of tolerance~\cite{traulsen(2003)pre}. They also extended this tag-based interactions to structured populations~\cite{traulsen(2004)pre} and well-mixed populations of finite size~\cite{traulsen(2007)plosone}. Tanimoto found that a two-dimensional tag space promotes cooperation more effectively than the one-variable tag system~\cite{tanimoto(2007)jtb}.

When it comes to spatial populations, Axelrod \emph{et al.} have found that cooperation can coevolve with tags in structured populations. Another far-reaching work~\cite{jansen(2006)nature} has probed the effects of beard chromodynamics on the evolution of cooperation, and revealed that loosening coupling between tag and strategy attenuates the oscillation of the population dynamics, reinforcing the beard color diversity and thus inducing higher levels of cooperation. Our work~\cite{wute(2013)jtb} has shown that adaptive tag switching can reinforce the coevolution of tag diversity and contingent cooperation even when tag switching is costly.

When strategy and tag mutate independently, a large body of work has addressed the evolution of cooperation under selection-mutation dynamics~\cite{antal(2009)pnas, tarnita(2009)pnas, tarnita(2011)pnas, zhang(2015)sr, ohtsuki(2006)prsb}. In Ref.~\cite{antal(2009)pnas}, the authors have considered a model with two strategies, contingent cooperation and defection. Each individual possesses a phenotype. Contingent cooperators just cooperate with these individuals of the same phenotype. By virtue of coalescent theory and perturbation theory, this work derived a very simple criteria, $(R-P)(1+\sqrt{3})>T-S$, for contingent cooperation to be selected. Using similar mathematical tools, Tarnita \emph{et al}. have considered the multiplexity of tags~\cite{tarnita(2009)pnas}. Each individual is affiliated with $n$ groups out of the total $m$ groups inhabitable. In this context, they gave the conditions under which cooperation can evolve. They have also applied this framework to study the competition of multiple strategies and give the terse condition for a specific strategy to be selected~\cite{tarnita(2011)pnas}.

Generally speaking, these models share two points in common. First, individuals are equally accessible to each phenotype (or tag) in a pre-assigned set of phenotypes available. And secondly, the interaction rate is binary. In the case of phenotypes being discrete, when sharing the same phenotype, two individuals interact. In the context of continuous tags, individuals just help those who differ with them in tag by no more than their tolerances.

Recent experimental studies have demonstrated that in terms of phenotype traits such as cooperativeness, resistance to antibiotic or competence, a high level of diversity can arise even in an isogenic population~\cite{acar(2008)ng, kussell(2005)genetics, wolf(2005)jtb, balaban(2004)science, ackermann(2008)nature, diard(2013)nature}. Individuals are capable of switching between phenotypes. By phenotype switching, the population can either optimize fitness~\cite{ackermann(2008)nature, diard(2013)nature}, or survive unpredictable environmental fluctuations~\cite{acar(2008)ng, kussell(2005)genetics, wolf(2005)jtb}, or preserve some properties~\cite{balaban(2004)science}.
The implications of these observations are threefold: individuals reserve redundant phenotypes; individuals are endowed with ability to switch phenotypes; fitness of phenotypes varies with environments. Meanwhile these studies have mainly concentrated on the species-environment systems, and delved into the importance of phenotype switching and its diversity for viability of organisms. Yet direct interactions between individuals are left largely unconsidered~\cite{acar(2008)ng, kussell(2005)genetics, wolf(2005)jtb, balaban(2004)science, ackermann(2008)nature, diard(2013)nature}.

Furthermore, although previous studies on phenotype similarity based cooperation~\cite{riolo(2001)nature, masuda(2015)pr, roberts(2002)nature, riolo(2002)nature, traulsen(2003)pre, traulsen(2004)pre, traulsen(2007)plosone, wute(2013)jtb, jansen(2006)nature, antal(2009)pnas, tarnita(2009)pnas, tarnita(2011)pnas, mcavity(2013)jtb} have considered direct interactions between individuals, these models have not dealt with difference in expressing phenotypes, leading to the binary interaction rate. A natural question arises what the evolutionary dynamics would be when individuals differ in ability of expressing phenotypes? We shall answer this question using a model integrating the diversity of phenotype expression and individual-individual interactions. Instead of uniformly expressing one of the potentially expressible phenotypes, our present model allows individuals each to express a subset of the potentially expressible phenotypes costly. The subset may vary with individual and evolves over time. This diversity in phenotype expression necessarily induces diversity in similarity. Unlike the zero-or-one interaction rate, two players met play the Prisoner's Dilemma game with likelihood positively relying on how many phenotypes they share of those they have expressed. The higher degree of the similarity between two individuals, the more likely they interact. In fact, several but not many studies have dedicated to investigating the effects of stochastic interactions~\cite{chen(2008)pre, traulsen(2007)jtb} on the evolution of cooperation. Traulsen \emph{et al}. have considered the stochasticity of interactions between neighbored individuals~\cite{traulsen(2007)jtb}. Chen \emph{et al}. have introduced the stochastic interactions into the spatial Prisoner's Dilemma game. In these studies~\cite{chen(2008)pre, traulsen(2007)jtb}, the interaction stochasticity is pre-assigned and thus does not evolve.

\section{Model}

Consider a well-mixed population of finite size $(=N)$. These $N$ individuals compete to survive. Reproduction is asexually and subject to mutation.
Each individual $i$ is characterized by a triplet $(s_i, K_i, G_i)$. The first entry $s_i$ denotes $i$'s strategy. In the Prisoner's Dilemma game, $s_i$ can be either cooperation or defection. For the sake of calculation, we let $s_i=1$ if $i$ is a cooperator and $s_i=0$ if $i$ is a defector. $G_i$ is the potentially expressible phenotypes individual $i$ carries. For we are mainly concerned with the diversity of phenotype expression, we assume $G_i$ is a constant, say $G_i\equiv G$. $K_i$ represents the phenotypes individual $i$ has actually expressed.

Each individual randomly expresses a number of the potentially expressible phenotypes. This means $K_i\subseteq G$. As the expression is random, even if two individuals express the same number of phenotypes, the specific phenotypes expressed can be different. Only phenotypes expressed are observable. Instead of assuming the zero-or-one interaction rate, we diversify the rate of interaction by associating it with the phenotypes shared in common. When two individuals $i$ and $j$ meet, the probability they play the game is dependent on how many identical phenotypes they possess of all the phenotypes expressed. Obviously, when both individuals $i$ and $j$ have expressed all phenotypes, they would interact with probability $1$.  When both share no phenotype of all those expressed, no interaction would happen. The larger number of phenotypes they share, the more likely they interact. Therefore, we introduce a function $r(K_i, K_j)$ to denote the interaction rate. For simplicity we first consider the linear interaction rate. In this linear mode, the probability that two individuals interact is proportional to the phenotypes they share (see figure 1).

At the same time, individuals need to bear the cost of expressing phenotypes. The cost is assumed to be proportional to the number of phenotypes expressed, that is, $\kappa_i(K_i)=\theta \cdot K_i$. Here we choose the simplest possible cost function of expressing phenotype. Therefore, the payoff consists of two parts: the payoff resulting from game interactions and the cost of expressing phenotypes. The net payoff $\pi_i$ determines the reproductive success of individual $i$. Here the fitness is an exponential function of payoff, say $f_i=e^{\beta \pi_i}$, where $\beta$ is the intensity of selection, specifying the contribution of the game to fitness.

The evolutionary updating is represented by a frequency-dependent Moran process. At each time step, each individual interacts with all other $N-1$ individuals depending on their expressed phenotypes and accrues payoffs. Payoff is mapped into fitness. Then an individual is chosen to reproduce an offspring with probability proportional to its fitness. Following birth, a randomly selected individual in the population is assigned to die and replaced. The population size thus remains constant throughout the evolution. Reproduction is subject to mutation. With probability $\mu$, the offspring with equal probability adopts one of the two behavioral strategies and also randomly expresses a number, say $K^{'}_i$, of phenotypes at a cost $\theta K^{'}_i$.


\section{Results}
We start with presenting the pairwise invasion dynamics, which shall help better understand the full population dynamics. When mutant defectors attempt to invade resident defectors, game interactions, whether or not they really happen, bring about nothing to payoff. This is due to the fact that defector-defector interaction generates no benefit to and charges no cost of the defectors involved. Payoff just consists of the cost of expressing phenotypes. The more phenotypes they express, the higher cost defectors pay. As a consequence, those defectors who express too many phenotypes place themselves in a least competent position. This explains why the plot peaks at the bottom right corner and unfolds downward towards the top left corner (see figure 2d).

In fact, this property can be rigorously corroborated. Suppose that the invading defectors express $K_Y$ phenotypes and the invaded defectors express $K_X$ phenotypes. Independent of the population composition and the phenotypes actually expressed, the fitness is $e^{-\beta \theta K_Y}$ and $e^{-\beta \theta K_X}$, for an invader and a resident, respectively. Denote by $\rho_{X\rightarrow Y}^{k}$ the fixation probability that a single mutant $Y$ takes over the resident population $X$ when $X$ and $Y$ share $k$ common phenotypes among their expressed phenotypes. We can readily obtain the fixation probability as $\rho_{X\rightarrow Y}^{k}=\frac{1-e^{\beta \theta (K_Y-K_X)}}{1-e^{\beta \theta N(K_Y-K_X)}}$ for all $k$ possibly allowed. Using Equation $(1)$ as presented in Methods Section, we can get the transition rate of the population moving from state $(X, K_X)$ to the state $(Y, K_Y)$ as $Q(X,Y; K_X, K_Y)=\frac{1-e^{\beta \theta (K_Y-K_X)}}{1-e^{\beta \theta N(K_Y-K_X)}}\cdot \sum_{k} 1$. It can be easily verified that the transition rate $Q(X,Y; K_X, K_Y)$ decreases with $K_Y$ and increases with $K_X$, respectively.

The pairwise dynamics exhibit similar cascading property when mutant cooperators attempt to invade resident defectors (see figure 2b). Obviously, expressing more phenotypes affects cooperators' evolutionary fate in three fronts. It undoubtedly reinforces reciprocity between cooperators themselves. It increases cost of expressing these phenotypes. In the third place, it also gets more likely for defectors to chase after these cooperators by expressing more phenotypes, and thus raising the rate of interaction between cooperators and defectors. Indeed, when resident defectors express a very few phenotypes, they have very little chance to exploit cooperators. This leaves more time space for mutant cooperators to reach the invasion barrier. As $K_X$ increases, the third effect becomes more conspicuous though invaded defectors also incur a bit higher cost of expressing phenotypes. These two negative effects offset and then overwhelms the reinforced reciprocity of mutant cooperators, resulting from incremental $K_Y$. In other words, when resident defectors have already expressed many phenotypes, mutant cooperators are unlikely to increase the chance of invading by expressing more phenotypes. These are reasons why the black line, corresponding to $1/N$ fixation probability, rises approximately linearly for small $K_X$, and gradually gets flattened as $K_X$ further increases.

Interesting scenarios are observed when mutant defectors attempt to invade resident cooperators (see figure 2c). Only when both mutation defectors and resident cooperators express a very large number of phenotypes, these defectors can take over the whole population with probability higher than $1/N$. This observation has an important implication. Whenever resident cooperators express so many phenotypes, defectors can expand their expressed phenotypes to exploit cooperators more severely. As it happens, payoffs for defectors due to the defector-cooperator interactions not only offset the cost of expressing more phenotypes but also put defectors in an advantageous position. This observation vanishes when resident cooperators express a very few phenotypes, suggesting that cooperators can dodge defectors' exploitation by unilaterally constraining phenotype expression.

The invasion dynamics show a saddle shape along the line directing from $(1,1)$ to $(20, 20)$ when mutant cooperators compete with resident cooperators (see figure 2a). No matter how many phenotypes, say $K_X$, resident cooperators express, mutant cooperators that express phenotypes of a number $K_Y$ most approximating $K_X$ are most likely to invade resident cooperators, though the fixation is still less than $\frac{1}{N}$. Reasons for this property vary as $K_X$ differs. For small $K_X$, when $K_Y$ is larger than $K_X$, the cost of expressing more phenotypes cannot be compensated by combined effects of sucking resident cooperators more and weak reciprocity between mutant cooperators. When $K_Y$ is less than $K_X$, mutant cooperators can hardly mutually breed, and thus being eclipsed by resident cooperators who always interact with each other albeit at a low rate. For large $K_X$, small $K_Y$ would prevent mutant cooperators from exploiting resident cooperators. Moreover, reciprocity between resident cooperators is so strong that mutant cooperators are least likely to invade. Even for $K_X$ in between, mutant cooperators cannot raise the fixation probability by expressing either more or less phenotypes than $K_X$.

With the pairwise invasion dynamics scrutinized, we are now able to address the full population dynamics. Consider the competition of an arbitrary number of strains $(=2G)$ in a finite-size population. The population is well-mixed. In the absence of mutation, the population dynamics inevitably end up in monomorphic state due to the stochastic nature of the evolutionary dynamics and the update rule. In the limit of small mutation, there are at most two strains competing in the population at the same time. Before next mutation occurs, the invaders either successfully take over the whole population or are wiped out. For this reason, the population dynamics can be approximated by an embedded Markov
chain of size $2G$, with each homogeneous state corresponding to one possible state of
the population associated with a strategy and a given number of phenotypes expressed. The transition rates between states are given as Equation $(1)$ in Methods Section.
The stationary distribution of this Markov chain characterizes the fraction of time the population spends in each of these $2G$ homogeneous states, and can be analytically computed.

Figure 3 presents the equilibrium level of these $2G$ strains as the parameter $\theta$ varies. Generally speaking, the population dynamics can be divided into three classes. For very small $\theta$, fraction of defective strains monotonically increases as phenotypes expressed grow (see figures 3a and b). When it comes to cooperative strains, the distribution of fractions exhibits a $U$-shape curve. It should be noted that the total level of defective strains is significantly higher than that of cooperative strains. As $\theta$ rises, for both cooperative strains and defective strains, those who express a very large number of phenotypes are depressed, while fractions of those expressing a very few phenotypes tilt upward (see figures 3c and d). For $\theta$ is as high as $0.10$, both distributions for defective strains and cooperative strains exhibit $U$-shapes, respectively. At this time, the total level of cooperative strains is comparable to that of defective strains. Further raising $\theta$ induces monotonically decreasing fraction distributions of both cooperative strains and defective strains (see figures 3d, e and f). That is, the more phenotypes expressed, the less fraction of corresponding strains. Cooperative strains enjoy remarkable advantage over defective strains in the evolutionary race. Explanations for these properties next follow.

For small $\theta$, the more phenotypes the defectors express, the more likely they exploit cooperators, the higher their fractions in the long run, a feature invariable as long as $\theta$ is lower than a certain value. For cooperative strains, a different picture emerges. In order to escape defectors' exploitation, cooperators have two choices: either to reduce phenotypes expressed or to reinforce reciprocity by expressing more phenotypes. Either choice turns out effective in defending defectors' invasion. Of the two, the later is more effective but still to a limited degree. Worst situation emerges when cooperators express a modest number of phenotypes. They can neither shun off interactions with defectors nor breed themselves strong enough to resist invasion of defectors. As a result, the fractions of these cooperative strains are lowest. A representative evolutionary process is invoked to present the core component of the population dynamics (see figure 4a).

As $\theta$ increases to moderate level such as $0.1$, advantages of defective strains expressing many phenotypes are depressed. It is because this advantage can be enjoyed only when competing cooperative strains also express a very large number of phenotypes. Recognizing this situation, cooperative strains can ward off invasion of defective strains by reducing expressing phenotypes. Once this happens, these cooperative strains always interact with each other with a constant small rate. Defective strains need to decipher which phenotypes cooperative strains have expressed from scratch. It takes them time. Hardly have they deciphered cooperative strains' phenotypes when they are already wiped out.

It makes defective strains harder to chase after and exploit cooperative strains as $\theta$ further increases. As mutation is uniform and unbiased, the evolutionary force uniquely determines the eventual fate of strains. Resident defective strains expressing a large number of phenotypes can be easily invaded by defective strains, or cooperative strains, when both just express not so many phenotypes. For resident defective strains
expressing a very few phenotypes, it may happen from time to time that they attempt to exploit cooperative strains by expressing more phenotypes, but as soon as they do so they will be pulled back. As a result, defective strains most of the time are entrenched in low levels with respect to phenotype expression. This no doubt constrains the ability of defective strains in deciphering and exploiting cooperative strains. For defective strains with low-level phenotype expression, they are outperformed by cooperative strains with similar level of phenotype expression as the later can enjoy the weak mutual breed almost certainly. The population dynamics are graphically illustrated by a typical process (see figure 4b).

Hitherto, our results have shown that cooperation coevolves with diversity of phenotype expression under a wider range of conditions and that expressing fewer phenotypes can best promote cooperation. It pays for cooperators to express a very few phenotypes, thereby improving their opportunity of establishing weak reciprocity, as these few randomly expressed phenotypes serve as secret handshakes and are difficult for defectors to discover and chase after. Trade-off between these two forces leads to the symbiosis of diversity of phenotype expression and the highest fraction of such cooperative strains. Our results also show that once defective strains dominate, the population dynamics are extremely unstable. In contrast, cooperative strains can prevail over the population over continuous periods of time, or long or short, but a prerequisite is that these cooperative strains must have randomly expressed a very few phenotypes. Thus a strong interdependence between the evolution of cooperation and diversity of phenotype expression is formed. Moreover, further investigation should aim to explore the forms of cost of expressing phenotypes and relationship between interaction rate and degree of similarity.

\section{Discussion}

Mounting efforts have been invested in exploring solutions to the evolution of cooperation~\cite{rakoff-nahoum(2016)nature, strassmann(2008)nature, wang(2010)tac, traulsen(2012)prsb, roca(2006)prl, wu(2012)pre, sasaki(2013)prsb, wu(2015)njp, zhang(2015)sr, ohtsuki(2006)prsb, perc(2006)njp, rand(2009)science, bednarik(2014)prsb}. Our study provides a possible path for establishing cooperation, in which the evolved diversity of phenotype expression plays a crucial role. Our study still comes into the domain of chromodynamics of cooperation~\cite{jansen(2006)nature, traulsen(2007)plosone}, but differs decisively from preceding studies on this topic~\cite{jansen(2006)nature, traulsen(2007)plosone}. In these studies, cooperators try to ward off defectors' exploitation through secret tags. When they run faster enough, cooperators dominate the population. A variety of ways can achieve this purpose~\cite{traulsen(2003)pre, traulsen(2004)pre, traulsen(2007)plosone, jansen(2006)nature, wute(2013)jtb}, such as prescribing high levels of phenotypic diversity~\cite{traulsen(2007)plosone}, weakening the coupling of tag and strategy~\cite{jansen(2006)nature}, introducing interaction stochasticity~\cite{chen(2008)pre}. The mechanism we proposed that interaction diversity induced by evolvable phenotype expression engenders similar effects as theirs~\cite{jansen(2006)nature}. In this sense, our mechanism is paralleled to them~\cite{jansen(2006)nature, traulsen(2007)plosone} and thus broadening `other mechanisms that can accomplish the same stabilizing effect' as the authors of Ref.~\cite{jansen(2006)nature} have suggested.

Moreover, our study is original in diversifying the expression of phenotypes. Previous studies dealing with interaction rates can be mainly divided into two classes. The first class~\cite{jansen(2006)nature, traulsen(2003)pre, traulsen(2004)pre, traulsen(2007)plosone} considers the zero-or-one interaction rate. When two individuals share the same (or similar enough) phenotype, they play game. Otherwise no interaction happens. In the second class~\cite{wute(2013)jtb, szolnoki(2016)pre, fu(2009)preb, pacheco(2009)prl, wu(2009)epl} interactions are contingent on the outcomes of previous interactions. This in fact forms a feedback loop. An exception comes when the interaction stochasticity is considered~\cite{chen(2008)pre}, but the stochasticity itself does not evolve.
Our model introduces the diversity of interaction rate by associating the rate with the degree of similarity, which is measured by the number of the same expressed phenotypes between individuals. This diversity does not rely on the information of preceding interactions. Natural selection acts on the population totally on the individual level.

Many mechanisms have been proposed maintaining phenotypic diversity under natural selection, such as mutant games~\cite{huang(2012)nc}, sexual selection~\cite{fisher(1930)book}, coevolving host-parasite population~\cite{milinski(2006)book}, occasional recombinations of tags and of strategies~\cite{jansen(2006)nature}, and phenotype noise~\cite{acar(2008)ng, kussell(2005)genetics, balaban(2004)science, wu(2017)ploscb, solopova(2014)pnas}. Diversity of phenotype expression can add to these mechanisms. Instead of assigning the number of phenotypes to express, our model allows individuals to express an arbitrary number of the potentially expressible phenotypes. The phenotypes expressed are subject to evolutionary force. Phenotypic diversity is therefore an evolvable trait. The frequency-dependent evolutionary dynamics drive the population to equilibrate at low levels of phenotype expression, and thus providing an adaptive explanation for the phenotypic diversity.

Two key themes in evolutionary biology, the evolution of cooperation and the diversity of phenotype expression, are naturally combined into our model. Populations seek to optimize fitness in evolutionary processes, while survival of cooperators are constrained through competition with defectors. To overcome the survival threat, cooperators may evolve the system of phenotype expression diversity through which they can distinguish potential partners. This conjecture is of pertinent consequence to many observations in biological circles~\cite{sinervo(2006)pnas, solopova(2014)pnas,corl(2010)pnas}. \emph{Salmonella Typhimurium} can either express virulence, or not. The side-blotched male lizards exhibit diverse colors in throats~\cite{sinervo(2006)pnas, solopova(2014)pnas}. In these examples, individuals have no difference in terms of carrying the potentially expressible phenotypes. They can regulate expression of phenotypes according to competitors faced, or even fluctuating environments. Our model is abstracted from these biological examples and has implications for and awaits the confirmation of field experimental studies.

\section{Methods}

\subsection{Fixation probability}
We here briefly illustrate the general procedure for calculating the fixation probability. In the limit of small mutation rate, the population simultaneously admits at most two different types of individuals, say $A$ and $B$. Denote by $K_A$ and $K_B$ the phenotypes expressed by individuals $A$ and $B$, respectively. Denote by $s_A$ and $s_B$ the behavioral strategies of individuals $A$ and $B$, respectively. Denote by $i$ the number of $A$s in the population. The Moran process that describes the evolutionary race has two absorbing states: $i=0$ and $i=N$. When the population arrives at either of these two states, it would stay there forever. Denote by $\phi_i$ the fixation probability that the population is eventually absorbed into the state $i=N$ when starting with the state $i$.

With this preparing work, we can easily write down the expected payoffs for an individual $A$ and $B$ as $P_A$ and $P_B$, respectively.
\begin{eqnarray}
P_A=r(K_A, K_A)(b-c) s_A (i-1)+r(K_A, K_B)(bs_B-cs_A) (N-i)-\theta K_A\nonumber
\end{eqnarray}
\begin{eqnarray}
P_B=r(K_A, K_B)(bs_A-cs_B) i+r(K_B, K_B)(b-c) s_B(N-i-1)-\theta K_B \nonumber
\end{eqnarray}
The fitness for $A$ and $B$ is $f_A=e^{\beta P_A}$ and $f_B=e^{\beta P_B}$, respectively. We consider the linear interaction rate as $r(K_A, K_B)=\frac{k}{|G|}$, with $k$ being the number of identical phenotypes individuals $A$ and $B$ possess. $|G|$ is the number of potentially expressible phenotypes carried. In an updating event, the population can increase one, decrease one, remain unchanged in terms of the number of $A$s, and corresponding probabilities are $T_{i, i+1}=\frac{if_A}{if_A+(N-i)f_B}\cdot\frac{N-i}{N}$, $T_{i, i-1}=\frac{(N-i)f_B}{if_A+(N-i)f_B}\cdot\frac{i}{N}$, and $T_{i, i}=1-T_{i, i+1}-T_{i, i-1}$, respectively. Then we have
\begin{eqnarray}
\phi_i=T_{i, i+1} \phi_{i+1}+T_{i, i-1} \phi_{i-1}+T_{i, i} \phi_{i} \nonumber
\end{eqnarray}
Using the boundary conditions $ \phi_{0}=0$ and $ \phi_{N}=1$ we can obtain the fixation probability as
\begin{eqnarray}
\phi_{1}=\big(1+\sum_{l=1}^{N-1}
\prod_{k=1}^l\frac{T_{k,k-1}}{T_{k, k+1}}\big)^{-1}\nonumber
\end{eqnarray}
\subsection{Transition rate for pairwise competing strains}
We here compute the rate the population transits from state $X$ to $Y$, which means the probability that strain $Y$ as mutant invades and takes over the population of $X$ strain. Suppose strain $X$ expresses $K_X$ phenotypes and strain $Y$ expresses $K_Y$ phenotypes. We should always bear in mind that as phenotype expression is random, for a giver number of phenotypes to express, the actually expressed phenotypes can vary. Denote by $\rho_{X\rightarrow Y}^{k}$ the fixation probability that a single mutant $Y$ takes over the resident population $X$ when $X$ and $Y$ share $k$ common phenotypes among their expressed phenotypes.

Then the expected transition rate from state $X$ to $Y$, say $Q(X, Y; K_X, K_Y)$, is given by

\begin{eqnarray}
Q(X, Y; K_X, K_Y)&=&\sum_{k=max\{0,K_X+K_Y-G\}}^{min\{K_X, K_Y\}}\frac{\Big(\begin{array}{c}K_x\\k\end{array}\Big)\Big(\begin{array}{c}G-K_x\\K_y-k\end{array}\Big)}{\Big(\begin{array}{c}G\\K_y\end{array}\Big)}\rho_{X\rightarrow Y}^{k}
\end{eqnarray}
Here $min\{x, y\}$ means the minimal of $x$ and $y$ and $max\{x, y\}$ means the maximal of $x$ and $y$, and $\Big(\begin{array}{c}K_x\\k\end{array}\Big)$ means the combination number of choosing $k$ from $K_x$. Some explanations on the lower and upper boundary of the summing are necessary. Whenever $K_X$ and $K_Y$ is no more than $G$, the expressed phenotypes strains $X$ and $Y$ share can be $0$. Whenever $K_X$ and $K_Y$ is larger than $G$, the expressed phenotypes strain $X$ and $Y$ share can be at least $K_X+K_Y-G$. The expressed phenotypes strain $X$ and $Y$ share can be at most $min\{K_X, K_Y\}$. This situation occurs when both strain $X$ and $Y$ happen to express phenotypes including these $min\{K_X, K_Y\}$ phenotypes. We can use Equation $(1)$ to analytically derive the transition rates between different population states in the limit of rare mutations and for any intensity of selection $\beta$.

\subsection{Stationary distribution} As the cost of expressing phenotype linearly increases with the number of phenotypes expressed, individuals expressing too many phenotypes would be easily invaded by those who express not so many phenotypes. In the long run, their fractions would be negligible. It is therefore reasonable to bound the number of potentially expressible phenotypes by a number $|G|$. In the limit of rare mutations, the population will most of the time stay in one of these $2G$ homogeneous states. In other words, there are simultaneously at most two differing strains present in the population before next mutation occurs. Therefore, the population dynamics of $2G$ strains can be well approximated by an embedded Markov chain between these $G$ full defective states and these $G$ full cooperative states. For convenience's sake, we label cooperative strains with even numbers $2K_C$, and defective strains with odd numbers,
$2K_D-1$, for $1\leq K_C \leq G$ and $1 \leq K_D \leq G$. For strain $X$ expressing $K_X$ phenotypes and strain $Y$ expressing $K_Y$ phenotypes, the expected transition rate from state $X$ to state $Y$ is $Q(X, Y; K_X, K_Y)$ as shown by Equation $(1)$. We can then easily get the transition matrix $A$ with dimension $2G \times 2G$. The $ij$th entry of matrix $A$ is $Q(i, j; K_i, K_j)$ for $i\neq j$, and the $ii$th entry is one minus the sum of all other entries in the $i$th row. It is worth noting that we have analytically derived the transition rates between any two competing strains and thus the transition matrix. The normalized left eigenvector associated with the eigenvalue $1$ of the transition matrix $A$ provides the stationary distribution of these $2G$ states. The overall cooperative level can be obtained by summing all the elements with even indices in normalized eigenvector~\cite{fudenberg(2006)jet, young(1993)e}.

\section*{Acknowledgements}
Financial support from NSFC (61751301 and 61533001) is gratefully acknowledged. Te Wu is also supported by the Fundamental Research Funds for Central Universities, Xidian University (JB180413).


\section*{Data, code and materials}
Codes (C++ and Matlab) and data that can be used to replicate the results in the paper are available upon request.


\clearpage
\begin{figure}[ht]
\centering
\caption{Random phenotype expression leads to interaction diversity. All individuals carry the same number, in this schematic illustration, say $G=5$, of potentially expressible phenotypes and their ability varies in expressing phenotypes. Each individual expresses a certain number of phenotypes at a cost proportional to the number. Phenotypes expressed take a toll on interaction rate. In panel $a$, individuals G1 and G2 each carry five expressible phenotypes. Individual $G_1$ expresses three phenotypes, say, Blue, Yellow and Green. G2 also expresses three phenotypes, say, Blue, Red and Green. Gray sections are unexpressed phenotypes and invisible. As the specific phenotypes vary in position, G1 and G2 just share one phenotype of those expressed. They would play the Prisoner's Dilemma game with probability $1/5$. In panel b, individuals G3 and G4 share three phenotypes in common of those expressed, they thus interact with probability $3/5$. The more phenotypes one expresses, the more likely he can interact with others yet the higher cost he shall bear on expressing these phenotypes.}
\label{xx}
\end{figure}

\clearpage
\begin{figure}[ht]
\centering
\caption{Pairwise invasion dynamics. Transition rate means the probability that the population moves from an invaded state to an
invading state. The capital letter, $C$ or $D$, along the $Y$-axis, denotes the mutant's behavioral strategy, while the one along the $X$-axis
denotes the residents' behavioral strategy. The coordinate value denotes the number of phenotypes expressed. Parameters: $N = 20$, $b=1.0$, $c=0.3$, $\beta = 0.1$, $\theta = 0.30$, $G=20$.}
\label{xx}
\end{figure}

\clearpage
\begin{figure}[ht]
\centering
\caption{Fraction of these $2G$ strains in stationary state. The bars are obtained by solving the eigenvector of the $2G \times 2G$ transition matrix. Blue denotes cooperative strain, and red defective strain. The abscissa value represents the number of phenotypes expressed. The evolutionary process is characterized by the Moran process, fully depicted in the main text. Parameters:  $N = 20$, $b=1.0$, $c=0.3$, $\beta = 0.1$, $\theta = 0.30$, $G=20$. In a, b, c, d, e, f, $\theta$ is $0$, $0.05$, $0.10$, $0.30$, $0.50$, $1.00$, correspondingly, and the overall cooperation level is $0.241$, $0.353$, $0.454$, $0.693$, $0.722$, and $0.667$, respectively.}
\label{xx}
\end{figure}

\clearpage
\begin{figure}[ht]
\centering
\caption{Time evolution of the competition between cooperative strains and defective strains. When cost of expressing phenotypes is low, the phenotype expression is most of the time stabilized at quite a high level. In this situation, interactions happen frequently. The evolutionary fate of strains is mainly determined by the game outcomes. Defective strains with high levels of phenotype expression dominate the evolutionary process. When cost of expressing phenotypes is high, such a high cost get hardly compensated by the payoffs resulting from game interactions. At this time, strains expressing fewer phenotypes are destined to be selected, which concomitantly takes defective strains more time to chase after and exploit cooperative strains. Cooperative strains, on the contrary, always interact with each other, though at a very slow rate. As a result, cooperative strains expressing a very few phenotypes overwhelm the evolutionary process. Parameters: $N = 20$, $b=1.0$, $c=0.3$, $\beta = 0.1$, $\theta = 0.30$, $G=20$, $\mu=6\times 10^{-5}$.}
\label{xx}
\end{figure}

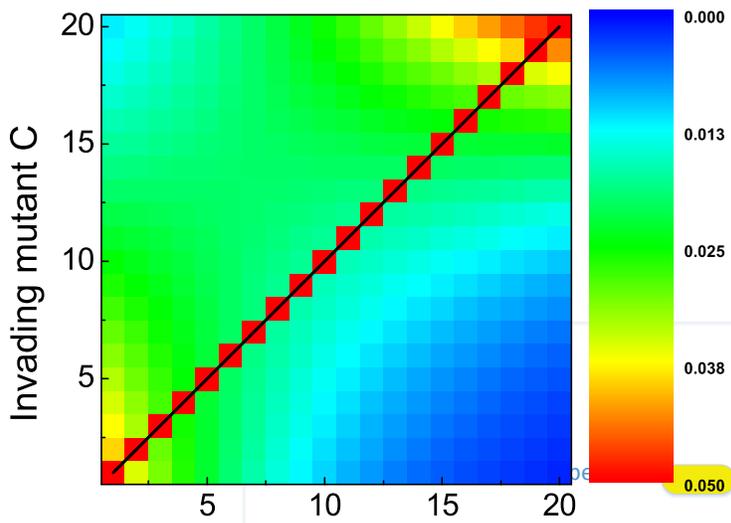
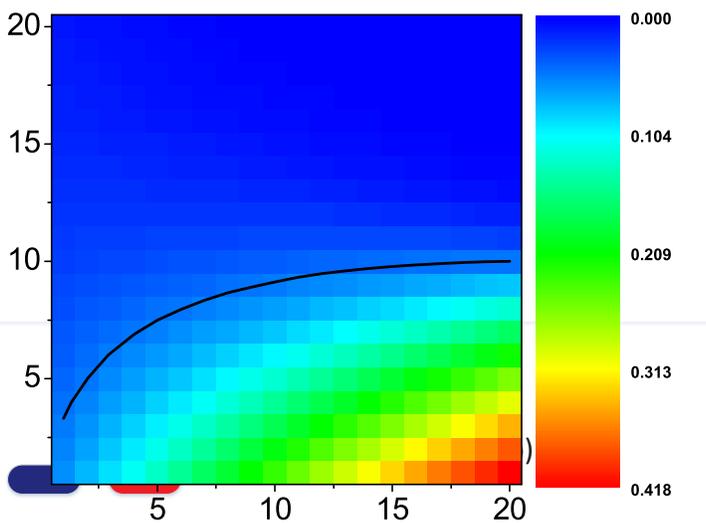
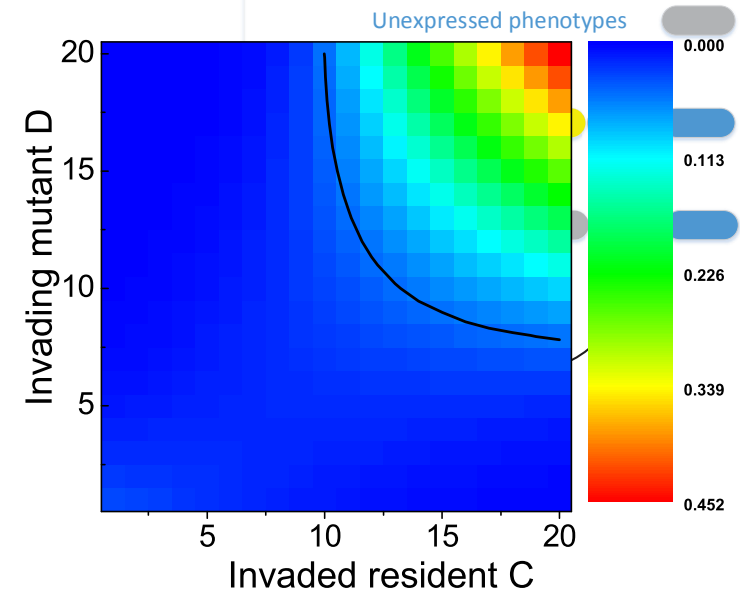
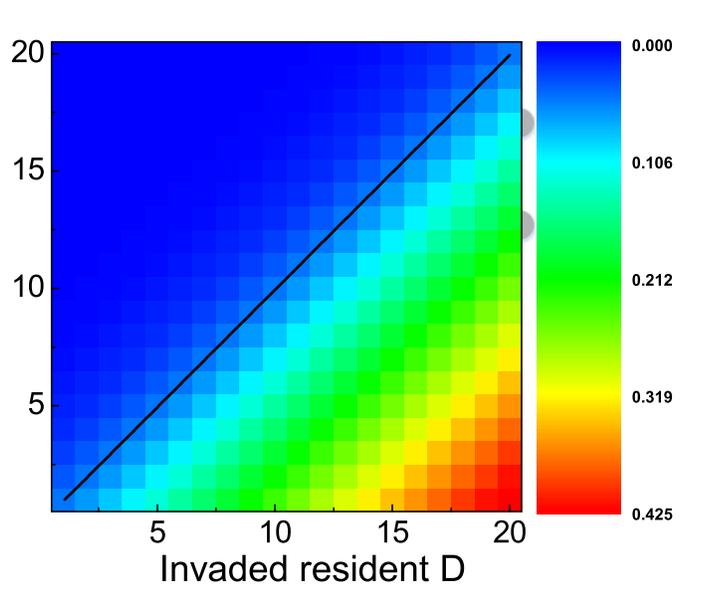
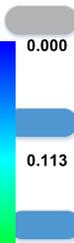
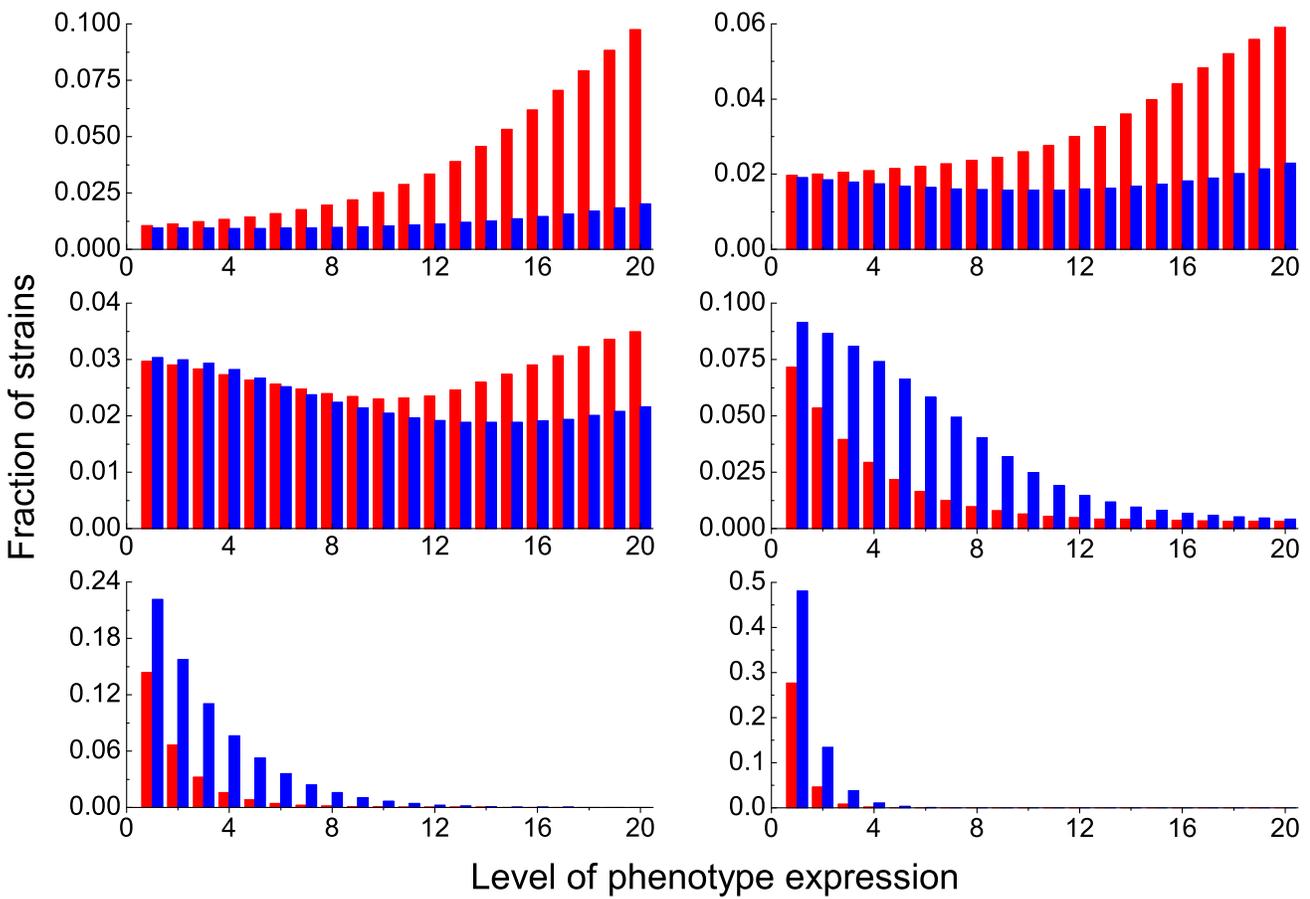
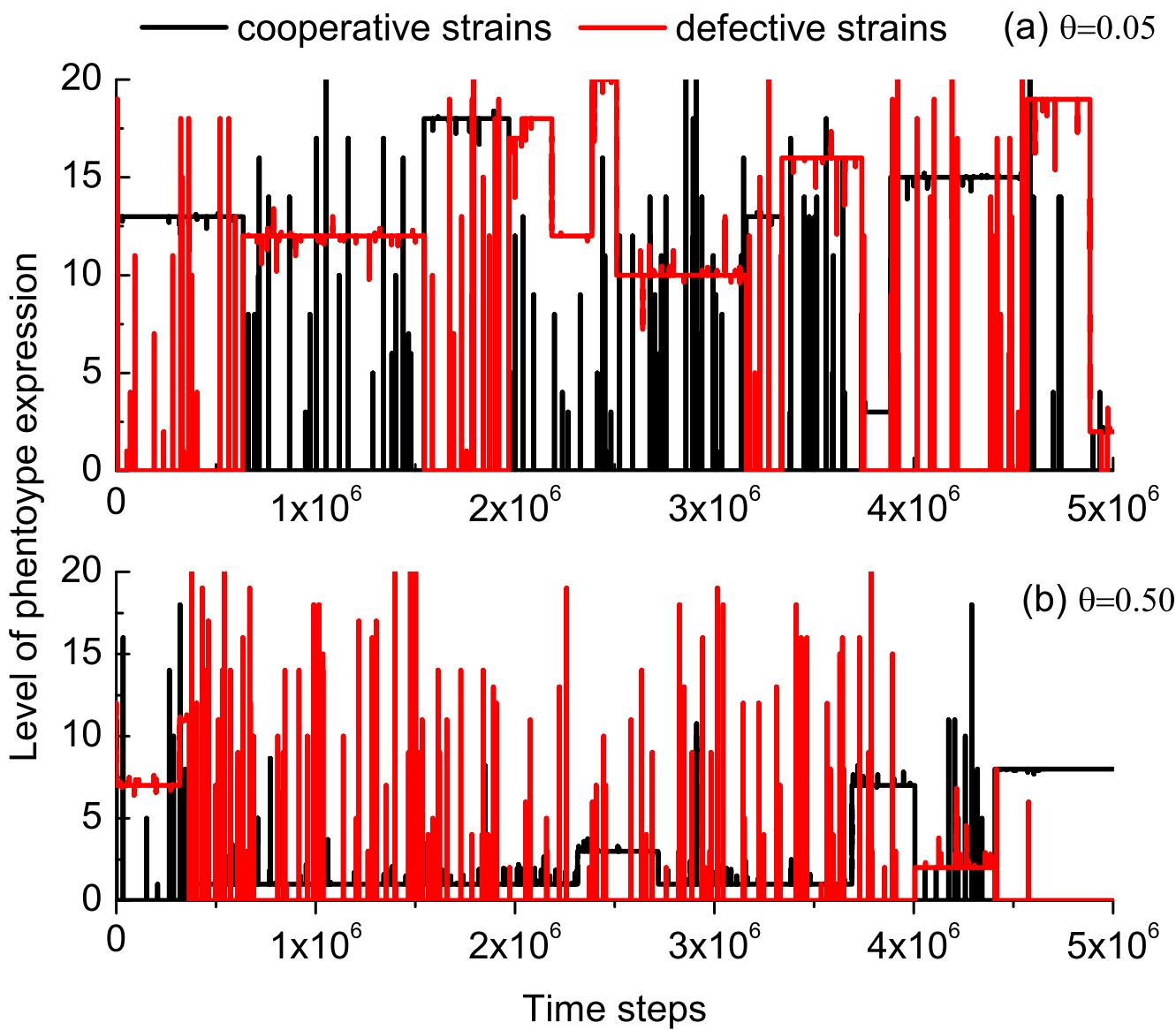
\end{document}